\documentclass[floatfix,showpacs,amsmath,amsfonts,amssymb,aps,twocolumn,
superscriptaddress,prb,10pt]{revtex4-1}
\usepackage[pdftex]{graphicx}
\usepackage[eulergreek]{sansmath}
\usepackage[caption=false,position=top]{subfig}
\usepackage{array}
\usepackage{dcolumn}

\newcolumntype{.}{D{.}{.}{-1}}

\newcommand{\rij}[0]{r_{ij}}
\newcommand{\vect}[1]{\mathbf{#1}}
\newcommand{\vecrij}[0]{\vect{r}_{ij}}

\newcommand{\expe}[0]{\mathrm{e}}

\newcommand{\figref}[1]{Fig.~\ref{#1}}

\newcommand{\neweqnline}{\nonumber\\}
\newcommand{\secref}[1]{Section~\ref{#1}}
\newcommand{\eqnref}[1]{Eq.~(\ref{#1})}
\newcommand{\tabref}[1]{Table~\ref{#1}}
\newcommand{\vecgrk}[1]{\boldsymbol{#1}}

\begin{document}
\title{Jastrow correlation factor for periodic systems}

\author{T.M.~Whitehead}
\affiliation{Cavendish Laboratory, J.J. Thomson Avenue, Cambridge, CB3 0HE,
  United Kingdom}
\author{M.H.~Michael}
\affiliation{Cavendish Laboratory, J.J. Thomson Avenue, Cambridge, CB3 0HE,
  United Kingdom}
\author{G.J.~Conduit}
\affiliation{Cavendish Laboratory, J.J. Thomson Avenue, Cambridge, CB3 0HE,
  United Kingdom}
\date{\today}

\begin{abstract}
  We propose a Jastrow factor for electron-electron correlations that 
interpolates between the radial symmetry of the Coulomb interaction at short 
inter-particle distance and the space-group symmetry of the simulation cell at 
large separation.  The proposed Jastrow factor captures comparable levels of 
the correlation energy to current formalisms, is 40\% quicker to evaluate, and 
offers benefits in ease of use, as we demonstrate in quantum Monte Carlo 
simulations.
\end{abstract}

\maketitle

\section{Introduction}

Quantum Monte Carlo (QMC) is a prominent family of techniques for studying 
strong correlations in quantum many-body systems \cite{Foulkes01}.  In 
particular, the variational and diffusion Monte Carlo methods (VMC and DMC) are 
accurate tools for studying ground-state energies and expectation values.  Both 
methods are predicated on the use of a trial wavefunction, whose similarity to 
the true ground state determines the accuracy and efficiency of the 
calculations \cite{Reynolds82}.  It is therefore important to have access to a 
high fidelity trial wavefunction.

A common foundation for constructing a fermionic trial wavefunction is to begin 
with the Hartree--Fock wavefunction $\Psi_\mathrm{HF}=D_\uparrow D_\downarrow$, 
where $D_\uparrow$ ($D_\downarrow$) is a Slater determinant of single-electron 
states for the up (down) spin species.  The Slater determinants encode the 
fermionic antisymmetry of the trial wavefunction, ensuring Pauli exchange is 
satisfied, but do not include any effects of electron correlation.  To describe 
such correlations, we  modify the trial wavefunction to be of the 
Slater--Jastrow \cite{Jastrow55} form $\Psi=\expe^{J(\vect{R})} D_\uparrow 
D_\downarrow$, where $\expe^{J(\vect{R})}$ is a Jastrow factor that is a 
function of all the electron positions, $\vect{R}$.  For real $J(\vect{R})$ the 
Jastrow factor is positive definite, and hence does not modify the nodal 
structure of the Hartree--Fock wavefunction.

In order to allow the Jastrow factor to accurately describe the correlations in 
a particular system of interest, $J(\vect{R})$ depends on a number of 
variational parameters 
\cite{McMillan65,Williamson96,Drummond04,Rios12,Ortiz94,Ceperley78,
Guclu05,Astrakharchik07}.  These parameters can be optimized 
using the relatively 
inexpensive VMC method, and then the optimal trial wavefunction used as the 
starting point for a more accurate but more expensive DMC calculation.  In 
principle the DMC estimate of the energy depends only on the nodal surface of 
the trial wavefunction \cite{Anderson76}, but in practice a more accurate trial 
wavefunction with an optimized Jastrow factor allows the method to proceed more 
efficiently.

In this paper we consider Jastrow factors for infinite, periodic systems.  
These systems are amenable to numerical simulation through the use of finite 
simulation cells which are tessellated, with periodic boundary conditions, to 
fill all of space.  Jastrow factors in the literature tend to either respect 
the short range radial symmetry of the Coulomb interaction, or abide by the 
symmetry of the simulation cells, but not both 
\cite{Williamson96,Drummond04,Rios12,Ortiz94}. 
 Here we propose a Jastrow factor that interpolates between these symmetries; 
is easier to use than current Jastrow factors by virtue of having a single 
parameter that tunes its accuracy, as opposed to two such parameters for other 
Jastrow factors of similar accuracy; requires fewer variational parameters to 
reach comparable accuracy; and is 40\% quicker to evaluate than these current 
Jastrow factors.

All of our QMC simulations were performed using the {\sc casino} package 
\cite{Needs10}, and we use Hartree atomic units throughout this paper.  In 
\secref{sec:Jastrow_factor} we review common Jastrow factors from the 
literature, and then show how our proposed Jastrow factor fits into this 
hierarchy.  In \secref{sec:HEG} and \secref{sec:beryllium} we examine the 
accuracy and efficiency of the Jastrow factors in the homogeneous electron gas 
and crystalline beryllium, respectively, before drawing our conclusions in 
\secref{sec:Discussion}.

\section{Jastrow factor}
\label{sec:Jastrow_factor}

We are concerned with Jastrow factors that capture correlation between 
electrons, and hence include functions of electron-electron separation, 
\begin{align*}
J(\vect{R}) = \sum_{\substack{j> i\\\sigma,\tau\in\{\uparrow,\downarrow\}}} 
J_{\sigma\tau}(\vecrij),
\end{align*}
where $\vecrij = \vect{r}_i-\vect{r}_j$, the sum runs over all electrons 
labeled $i$, $j$, and we refer to $J_{\sigma\tau}(\vecrij)$ as a Jastrow 
function.  The
Jastrow function contains variational parameters that we optimize within a VMC
calculation to minimize the variance in the local energy \cite{Umrigar88}. 

There are some fundamental constraints on the form of the Jastrow function. 
Firstly, in order to retain the spin expectation value of the Hartree--Fock 
wavefunction $J_{\sigma\tau}(\vecrij)$ must be even under exchange of 
particles. Secondly, in order to avoid non-physical divergences in the 
local energy, 
$J_{\sigma\tau}(\vecrij)$ must be at least twice-differentiable everywhere 
except at particle coalescence ($\vecrij=\vecgrk{0}$).

However, at particle coalescence the Coulombic potential energy of two 
electrons diverges.  In order to retain a non-divergent local energy the 
kinetic energy therefore has to diverge in the opposite direction at particle 
coalescence.  This may be achieved by imposing the Kato cusp conditions 
\cite{Kato57} on the wavefunction, which may be expressed as
\begin{align*}
\left. \frac{\partial J_{\sigma\tau}}{\partial \rij} \right|_{\rij=0} = 
\Gamma_{\sigma\tau},
\end{align*}
giving spherically-symmetric behavior at short radius 
$J_{\sigma\tau}(\vecrij)=\Gamma_{\sigma\tau}r_{ij}+\ldots$, where, for 3D
systems, 
\mbox{$\Gamma_{\uparrow\uparrow}=\Gamma_{\downarrow\downarrow}=\frac{1}{4}$} 
and 
$\Gamma_{\uparrow\downarrow}=\Gamma_{\downarrow\uparrow}=\frac{1}{2}$.  The 
final constraint on the Jastrow factor is that, in periodic systems like those 
we consider here, $J_{\sigma\tau}(\vecrij)$ must satisfy periodic boundary 
conditions at the edge of the simulation cell in order to tessellate space.

Before presenting and testing our proposal for a Jastrow factor, we first 
review other Jastrow factors that are commonly used in the literature.  We 
organize the Jastrow factors by their symmetry, starting with a spherically 
symmetric function and then examining a Jastrow factor with the symmetry of the 
simulation cell before proposing our Jastrow factor that interpolates between 
these symmetries.

\subsection{Term with spherical symmetry}

The interaction between two isolated electrons is isotropic, and so it is 
reasonable to take the Jastrow factor as being spherically symmetric and purely 
a function of particle separation where two-body effects dominate, and 
especially at inter-particle separations shorter than the average 
nearest-neighbor separation in many-body systems.  However, the simulation 
cells used in numerical calculations are not spherically symmetric as they have 
to tessellate to fill 3D space.  Because of this requirement, and in order to 
limit the effect of otherwise infinite-ranged terms to within the simulation 
cell, radial terms in the Jastrow factor are cut off at a finite radius that is 
less than or equal to the Wigner--Seitz radius corresponding to the simulation 
cell.  This is implemented by including a term $(1-\rij/L_{\sigma\tau})^C 
\Theta (L_{\sigma\tau}-\rij)$ in the Jastrow function, which goes to zero at a 
radius $L_{\sigma\tau}$, with $C-1$ continuous derivatives.  We take $C=3$ in 
order to keep the local energy continuous at the cutoff radius 
\cite{Drummond04}.  $\Theta (L_{\sigma\tau}-\rij)$ is a Heaviside step 
function, which forces the Jastrow function to be zero everywhere beyond the 
radius $L_{\sigma\tau}$.

It has been found in the literature 
\cite{Drummond04,Rios12,AlHamdani14,Chen14,Gillan15} that a 
Taylor expansion in electron-electron separation captures the most important 
short-ranged isotropic inter-particle correlations, and so here we review that 
expansion.  Writing the Jastrow correlation function as a Taylor series around 
particle coalescence results in an expression 
\begin{align}
u_{\sigma\tau}(\rij) =& \left( \frac{L_{\sigma\tau}}{3}\left[ 
\alpha_{1,\sigma\tau} - \Gamma_{\sigma\tau} \right] + \sum_{m=1}^{N_\mathrm{u}} 
\alpha_{m,\sigma\tau} \rij^m\right) \times \neweqnline
&\times \left( 1-\rij/L_{\sigma\tau} \right)^3 \Theta\left(L_{\sigma\tau} - 
\rij \right),
\label{u_term}
\end{align}
which is referred to as a $u$ term \cite{Drummond04,Rios12}.  Here the 
$N_\mathrm{u}$ coefficients $\alpha_{m,\sigma\tau}$ are parameters that are 
optimized using VMC, and the cutoff length $L_{\sigma\tau}$ is also optimized 
variationally.  The term $L_{\sigma\tau}[ \alpha_{1,\sigma\tau} - 
\Gamma_{\sigma\tau} ]/3$ ensures that the Kato cusp conditions are satisfied.
Using a pseudopotential for the electron-electron 
interaction \cite{LloydWilliams15} would set $\Gamma_{\sigma\tau}=0$.

\begin{figure*}
 \subfloat[$u$ term]{
 \includegraphics[width=0.5\linewidth]{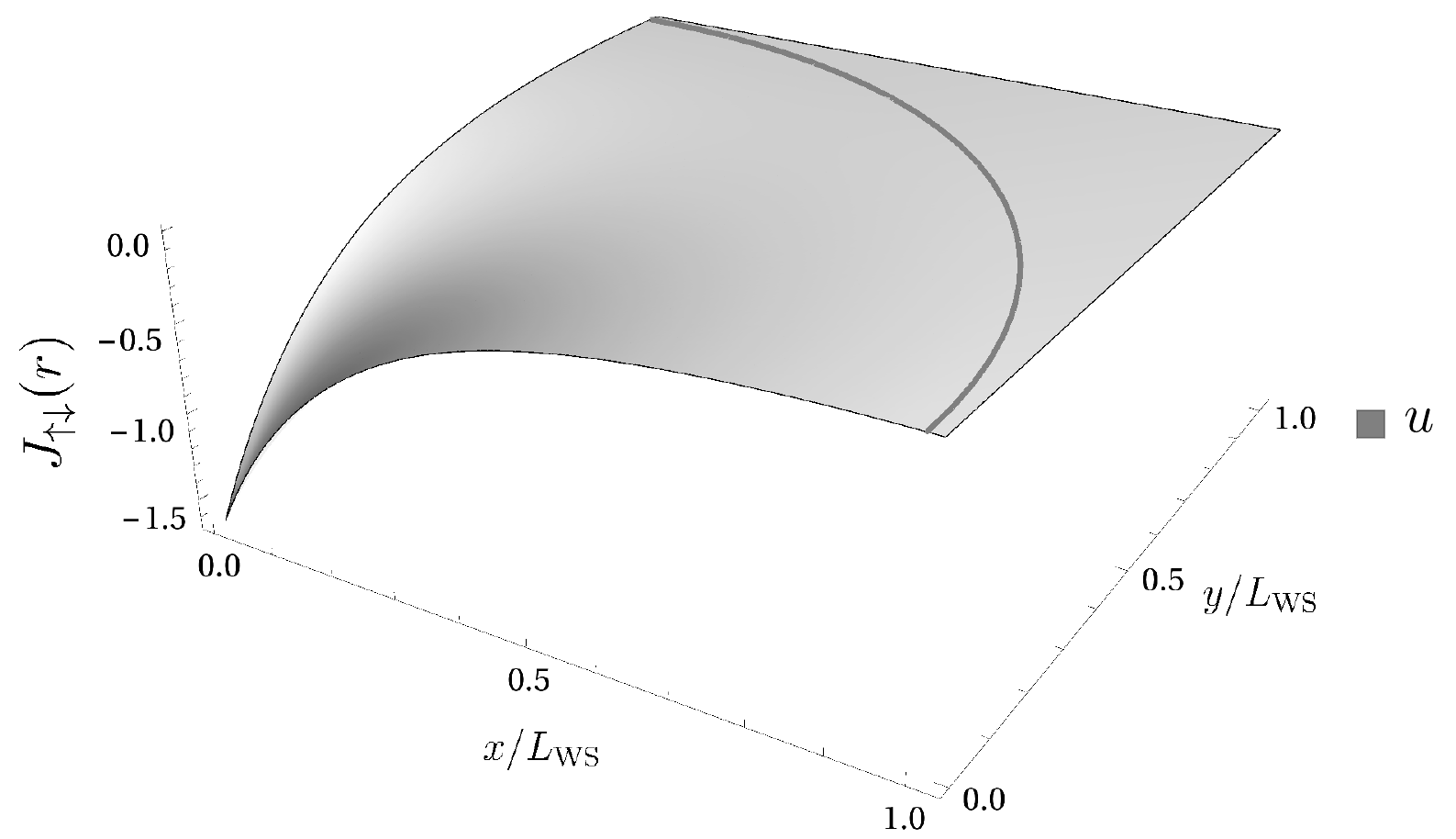}
 \label{fig:functions_u}
 }
 \subfloat[$p$ term]{
 \includegraphics[width=0.5\linewidth]{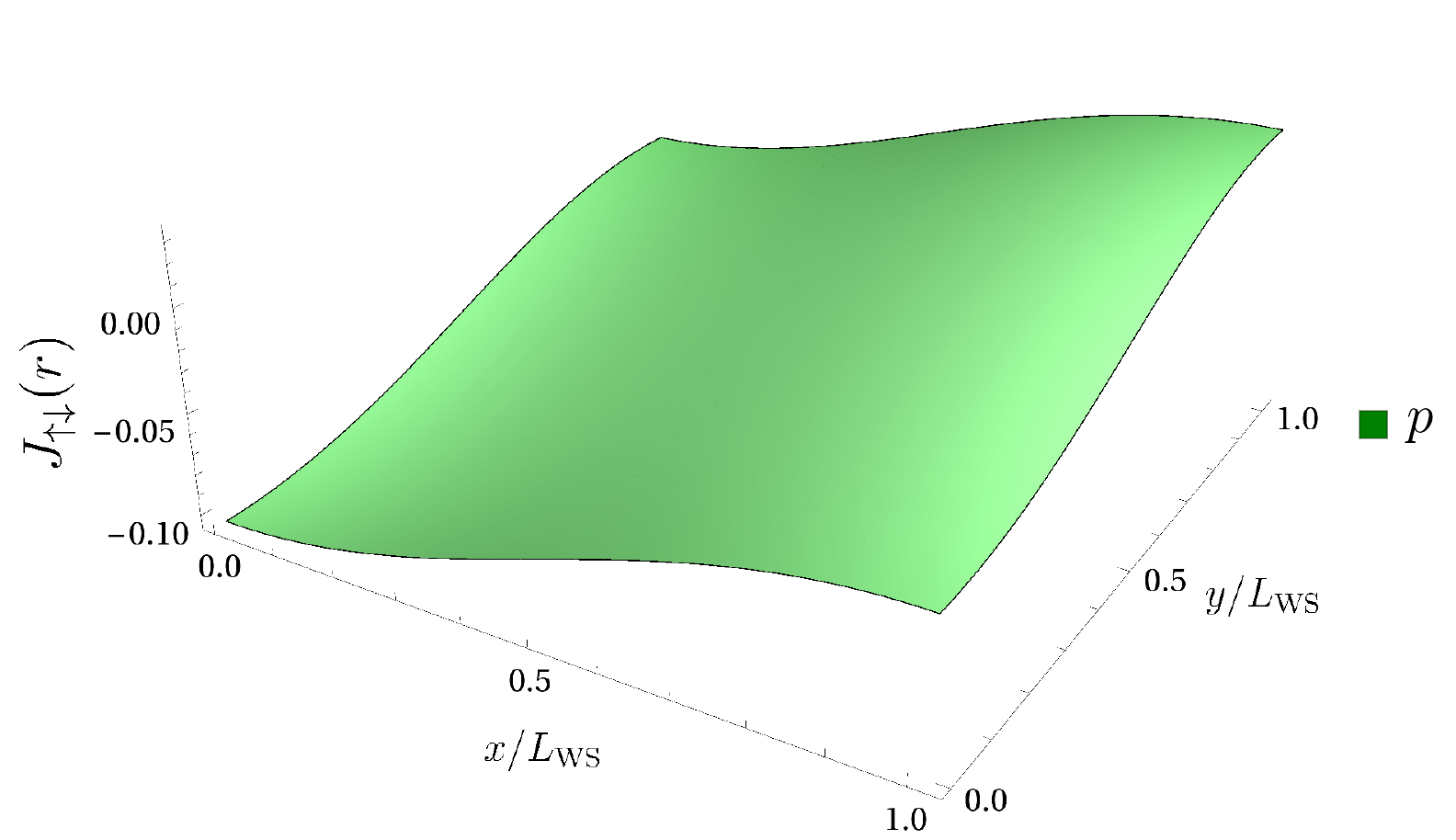}
 \label{fig:functions_p}
 }
\\
 \subfloat[$u$ and $u$ \& $p$ terms]{
 \includegraphics[width=0.5\linewidth]{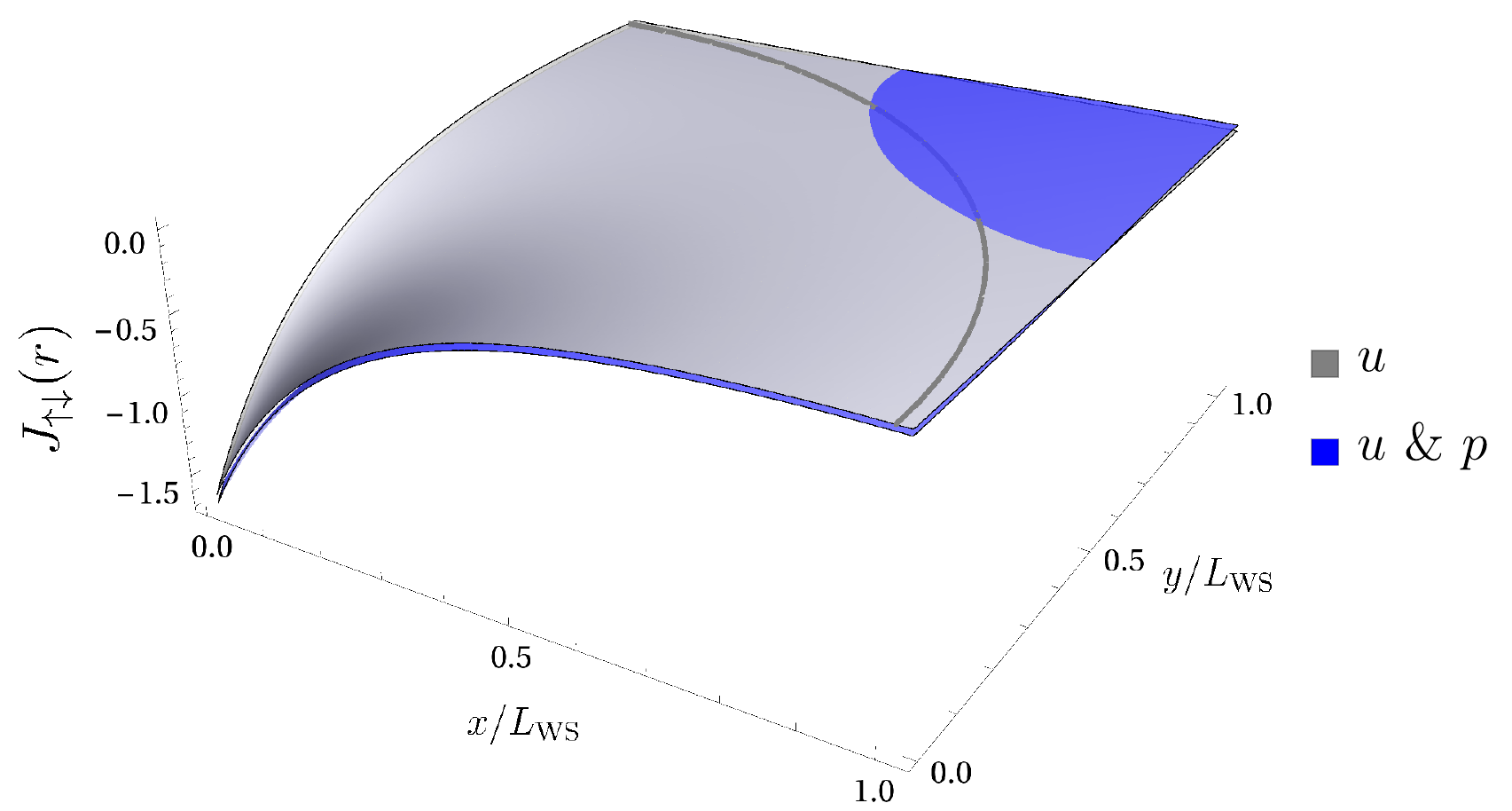}
 \label{fig:functions_u1p}
 }
 \subfloat[$u$ and $\nu$ terms]{
 \includegraphics[width=0.5\linewidth]{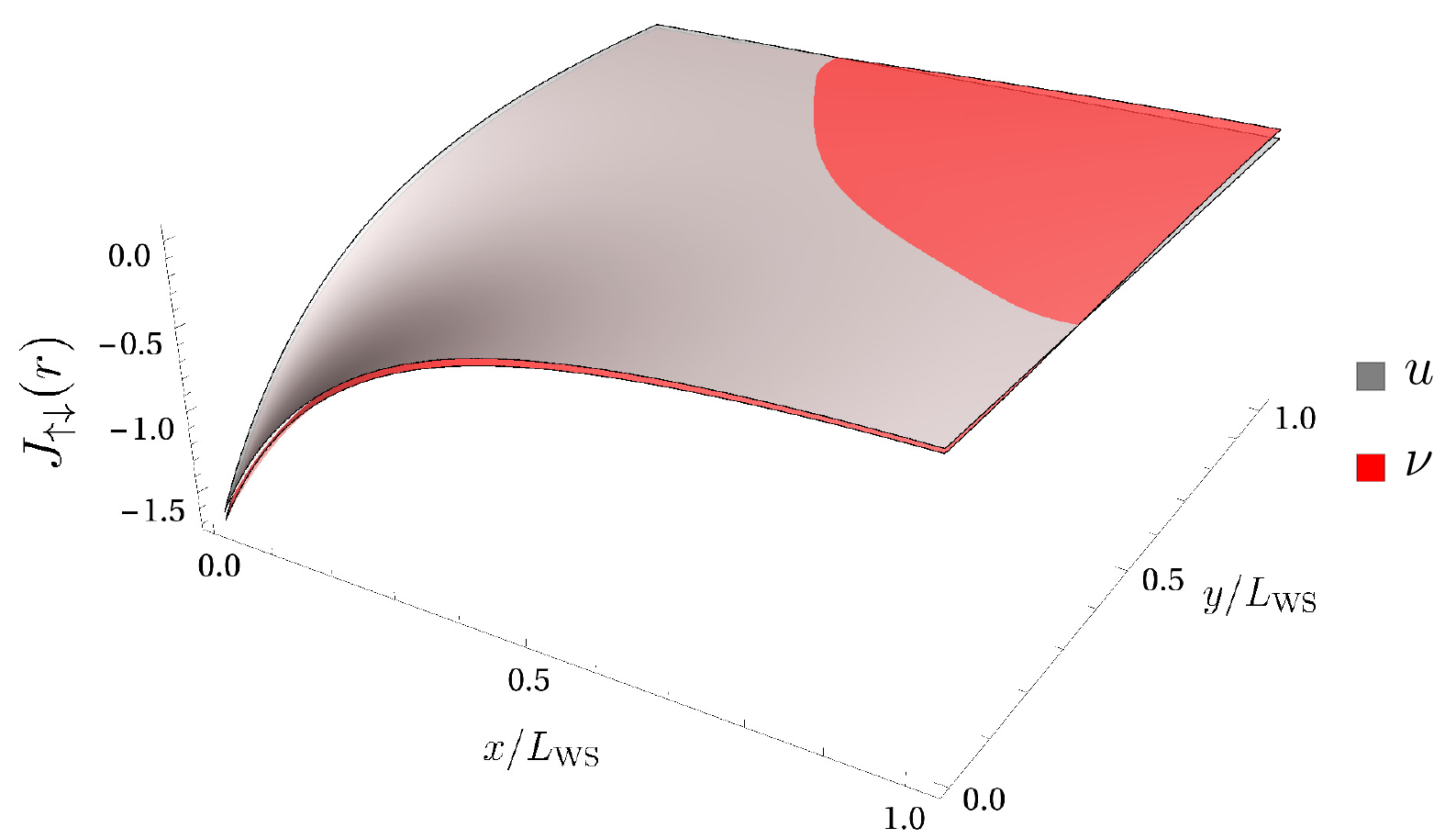}
 \label{fig:functions_nu}
 }
 \label{fig:jastrow_fns}
 \caption{(Color online)  The Jastrow functions discussed in the main text, 
showing the (a) $u$, (b) $p$, (c) $u$ (gray) and $u$ \& $p$ (blue), and (d) 
$u$ (gray) and $\nu$  (red)
functions. The $u$, $u$ \& $p$, and $\nu$ 
terms each have a total of five variational parameters, optimized using VMC in 
the homogeneous electron gas system.  The data are taken for an opposite-spin 
electron pair in the $z=0$ plane, with one particle at the origin, showing one 
quadrant of the simulation cell.  A gray arc indicates the $u$ term's cutoff 
radius, which is comparable to the Wigner--Seitz radius $L_\mathrm{WS}$ of the 
simulation cell.  We have subtracted a physically-irrelevant constant from the 
$\nu$ Jastrow function for clarity.
}
\end{figure*}

The $u$ term Jastrow function with parameters optimized for an homogeneous 
electron gas with $r_\mathrm{s}=4$ is shown in \figref{fig:functions_u}.  The 
short-range behavior of the $u$ term is linear to satisfy the Kato cusp 
condition, and then at large separation the cutoff function limits the range of 
the $u$ term to within the Wigner-Seitz radius of the simulation cell, shown as 
a gray arc in \figref{fig:functions_u}.  This not only limits the maximum range 
of the correlations that can be captured by the $u$ term, but also prevents it 
from capturing correlations in the corners of the simulation cell.

\subsection{Term with simulation cell symmetry}

One method to extend the ability of the Jastrow factor to capture correlations 
over the whole simulation cell is to use a form of Jastrow factor that innately 
has the space-group symmetry of the simulation cell. A simple but effective 
example of a Jastrow function that has such symmetry is a plane-wave basis, 
which also explicitly ensures periodicity of the Jastrow function.  The 
so-called $p$ term takes the form \cite{Drummond04,Rios12}
\begin{align}
p_{\sigma\tau}(\vecrij) = \sum_{\ell=1}^{N_\mathrm{p}} a_{\ell,\sigma\tau} 
\sum_{\vect{G}_\ell^+} \cos ( \vect{G}_\ell \cdot \vecrij ).
\label{p_term}
\end{align}
Here the $\{ \vect{G}_\ell \}$ are the reciprocal lattice vectors of the 
simulation cell that belong to the $\ell^\mathrm{th}$ star of vectors 
equivalent under the full symmetry group of the simulation cell, sorted by 
increasing size of $|\vect{G}_\ell|$ (and in periodic systems not including the 
trivial vector $\vecgrk{0}$); ``$+$'' means that if $\vect{G}_\ell$ is included 
in the sum, $-\vect{G}_\ell$ is excluded; and the $a_{\ell,\sigma\tau}$ are 
variational parameters, of which there are $N_\mathrm{p}$.

A $p$ term with $N_\mathrm{p}=1$ for the homogeneous electron gas is shown in 
\figref{fig:functions_p}.  The $p$ term exists over the whole simulation cell, 
including where the $u$ term is cut off to zero.  This means the $p$ term can 
capture correlations in the cell corners that the $u$ term misses.  However, 
the $p$ term does not tend to a radial form at short radius and so cannot 
satisfy the Kato cusp conditions at particle coalescence, meaning that on its 
own it does not make for an effective Jastrow factor.  A common approach in the 
literature 
\cite{Conduit09,Drummond13,Whitehead16,Chiesa06,Spink13,Keyserlingk13,
Maezono10,Mostaani15} is to combine both $u$ and $p$ terms to give a composite 
Jastrow function
\begin{align}
J_{\sigma\tau}(\vecrij)= u_{\sigma\tau}(\rij) + p_{\sigma\tau}(\vecrij),
\label{up_jastrow}
\end{align}
which uses the $u$ term to capture short-range correlations and the Kato cusp 
conditions, and the $p$ term to capture long-range correlations in the corners 
of the simulation cell.  We refer to such a combination as a $u$ \& $p$ term.

An example of this composite Jastrow function, with $N_\mathrm{u}=3$ and 
$N_\mathrm{p}=1$ and parameters optimized in an homogeneous electron gas, is 
shown in \figref{fig:functions_u1p}.  As expected, the behavior at short range 
is dominated by the $u$ term. Yet at large radius this Jastrow function has 
structure due to the $p$ term, including in the corner of the simulation cell 
outside the cutoff radius of the $u$ term, shown by the gray arc, which allows 
the composite $u$ \& $p$ term to capture longer-range correlations.  

However, this construction has several undesirable features that limit its 
effectiveness at capturing inter-particle correlations.  For a given amount of 
computing time to be spent optimizing the parameters in the Jastrow factor, a 
choice needs to be made of the relative number of $u$ and $p$ terms to be used. 
 We do not know a priori the optimal ratio of $N_\mathrm{u}$ to $N_\mathrm{p}$, 
and so must explore a two-dimensional parameter space to determine it.  A large 
proportion of the VMC calculation time is spent evaluating the Jastrow factor, 
and so it is important that the Jastrow factor is as simple as possible. But 
there is not equality of expense between the $u$ and $p$ terms, as sinusoidal 
$p$ terms are more expensive to calculate than polynomial $u$ terms, and the 
expense of a $p$ term also increases with the number of elements of the 
reciprocal lattice vector stars used to evaluate it. Higher-order stars 
generally contain more elements than lower-order ones, meaning high-order $p$ 
terms are even more expensive to calculate.  To further complicate the 
optimization of the $u$ \& $p$ term, although the $p$ term was intended to 
capture longer-range correlations, it does also exist at short radius; this 
means it interferes with the effect of higher-order contributions from the $u$ 
term.

One further problem with the form of Jastrow function given by 
\eqnref{up_jastrow} is that the cutoff length $L_{\sigma\tau}$ enters the 
expression non-linearly.  To optimize the cutoff length and other parameters we 
need to solve a multi-dimensional non-linear set of equations, which is a 
significantly more difficult problem than solving a multi-dimensional linear 
set of equations, where the full force of linear algebra may be applied to 
increase the efficiency of the process \cite{Drummond05}. 

We are interested in finding a form for the Jastrow factor that avoids these 
problems with the current method, by being a term with a single tuning 
parameter that determines the accuracy of the Jastrow factor, and which is also 
cheap to evaluate with linear coefficients.  At the same time the proposed term 
should reproduce the advantageous properties of the $u$ term, accurately 
capturing short-range correlations, and also the $p$ term, exhibiting the 
symmetry of the simulation cell at large inter-particle separation.

\subsection{\mbox{\boldmath$\nu$} term}
\label{sec:nu}

We propose a Jastrow factor that combines the properties and symmetries of the 
$u$ term at small radius with the properties and symmetries of the $p$ term at 
large separation.  The Jastrow function, referred to here as the $\nu$ term, is
\begin{align}
\nu_{\sigma\tau}(\vecrij) &= \sum_{n=1}^{N_\nu} c_{n,\sigma\tau} \left| 
f_\mathrm{x}^2(\vect{x}_{ij}) 
+f_\mathrm{y}^2(\vect{y}_{ij})+f_\mathrm{z}^2(\vect{z}_{ij}) \right|^{n/2}, 
\neweqnline
f_\mathrm{x}(\vect{x}) &= |\vect{x}| \left( 
1-\frac{|\vect{x}/L_\mathrm{x}|^3}{4} \right),
\label{eq:cubic_nu}
\end{align}
where the $N_\nu$ parameters $c_{n,\sigma\tau}$ are optimized using VMC, and 
the length $L_\mathrm{x}$ is the width of the simulation cell in the 
Cartesian $x$-direction.  In \secref{sec:gennu} below we generalize the $\nu$
term to non-cuboidal geometries.

At small radius, the function $f_\mathrm{x}(\vect{x})=|\vect{x}|+ 
O\left(|\vect{x}|^4\right)$, and so 
\mbox{$|f_\mathrm{x}^2(\vect{x})+f_\mathrm{y}^2(\vect{y})+f_\mathrm{z}^2(\vect{z
})|^{1/2}=r+O\left( r^4 \right)$}.  This has the correct spherical symmetry to 
describe short-range electron-electron correlations, so at short radius the 
Jastrow function $\nu_{\sigma\tau}(\vecrij) = \sum_{n=1}^{N_\nu} 
c_{n,\sigma\tau} \rij^n+\ldots$ consists of an expansion in electron-electron 
separation, similarly to the $u$ term.  This means the $\nu$ term will 
reproduce the ability of the $u$ term to capture short-ranged correlations and 
it is easy to satisfy the Kato cusp conditions by setting $c_{1,\sigma\tau} = 
\Gamma_{\sigma\tau}$.

The function $f(\vect{x})$ is symmetric under $\vect{x}\to -\vect{x}$, and 
automatically satisfies periodic boundary conditions at the edge of the 
simulation cell, with $f(L_\mathrm{x}\hat{\vect{x}})\neq 0$, 
$f'(L_\mathrm{x}\hat{\vect{x}})=0$, and $f''(L_\mathrm{x}\hat{\vect{x}})\neq 
0$: this is achieved through the use of the cubic power in the definition of 
$f$, chosen by analogy to the cutoff function in the $u$ term to distinguish 
long- and short-ranged components of the Jastrow function.  Importantly, the 
$f$ functions satisfying periodic boundary conditions means any function 
constructed from them, such as the $\nu$ term, will also correctly satisfy 
periodic boundary conditions. The scaling of the $f$ functions in the different 
Cartesian directions lends the $\nu$ term the symmetry of the simulation cell 
at large inter-particle separation, and allows the $\nu$ term to capture 
long-range correlations, similarly to the $p$ term.  Not requiring a cut-off 
function also means all the variational parameters enter the expression for 
$\nu_{\sigma\tau}(\vecrij)$ linearly, and so are easier to optimize than the 
equivalent number of variational parameters in the $u$ term \cite{Drummond05}.

The $\nu$ Jastrow function optimized for a homogeneous electron gas is shown in 
\figref{fig:functions_nu}, demonstrating that it has the same small-radius 
behavior as the $u$ term.  We can also see that the $\nu$ term still has 
structure in the corner of the simulation cell, similarly to the $u$ \& $p$ 
term, which allows it to capture long-range inter-particle correlations.  We 
will examine this similarity in more detail in a case study of the homogeneous 
electron gas in \secref{sec:HEG}.

Freedom to optimize the behavior of the Jastrow factor in the corners of the 
simulation cell also provides the freedom to change the kinetic energy of the 
wavefunction in the corners of the simulation cell, as 
$f_\mathrm{x}''(L_\mathrm{x}\hat{\vect{x}})\neq 0$.  This allows the $\nu$ 
Jastrow factor to more accurately respond to a finite and/or varying potential 
energy in the corners of the simulation cell. From a Thomas--Fermi perspective 
this provides the $\nu$ term with the freedom to counteract changes in the 
potential energy from interactions with kinetic energy in order to keep the 
total energy constant.

The $\nu$ term may also be adapted to systems other than the
3D ones considered here.  For 2D systems the $f_\mathrm{z}$ function may 
simply be omitted; or for slab geometries, with two directions periodic and 
one non-periodic, $f_\mathrm{z}$ should be replaced by a function
that reduces to $|\vect{z}|$ at short radius, for example 
$|\vect{z}|\expe^{-(|\vect{z}|/L_\mathrm{z})^2}$.

In order to demonstrate the advantages of the $\nu$ Jastrow factor, in the 
next two Sections we carry out simulations of the homogeneous electron gas and 
a crystalline solid.  We examine the accuracy, efficiency, and ease of use of 
the $\nu$ Jastrow factor, and compare it with other forms of Jastrow factor 
used in the literature.

\section{Homogeneous electron gas}
\label{sec:HEG}

For the first test of our Jastrow factor we examine the homogeneous electron 
gas (HEG).  This system has been widely studied using QMC 
\cite{Ceperley80,Zong02,Drummond04a,Spink13} and serves as an 
analogue for electrons in a conductor. As it does not contain any atoms it 
allows us to focus on the electron-electron Jastrow factor.  For simplicity we 
assume that the intra-species correlations for the up- and down-spin
electrons are 
identical, and so fix $J_{\uparrow\uparrow}=J_{\downarrow\downarrow}$ and 
$J_{\uparrow\downarrow}=J_{\downarrow\uparrow}$.

We examine a HEG with density parameter $r_\mathrm{s}=4$ in a cubic simulation 
cell subject to periodic boundary conditions, and use Slater determinants of 
plane wave orbitals.  We use a system of 57 up- and 57 down-spin electrons, and 
confirmed that the main results of this section were reproduced in systems of 
33 and 81 electrons per spin species and so are independent of system size.  We 
optimize all the Jastrow factors by minimizing the variance in the local energy 
\cite{Drummond05,Kent99}, and confirmed that minimizing the energy directly 
\cite{Umrigar07} gave similar results. All VMC 
simulations are run for $1\times10^6$ steps.  We then carry out DMC simulations 
to obtain a more accurate estimate for the energy within the fixed node 
approximation, $E_\mathrm{DMC}$, which corresponds to the use of a perfect 
Jastrow factor.  DMC simulations starting with different trial wavefunctions 
agree to within $5\times10^{-6}$ a.u. To measure the accuracy of the Jastrow 
factors, we evaluate the percentage of the DMC correlation energy missing from 
the VMC simulation,
\begin{align*}
\eta = \frac{E_\mathrm{VMC}-E_\mathrm{DMC}}{E_\mathrm{HF}-E_\mathrm{DMC}} 
\times 100 \%
\end{align*}
where the Hartree--Fock energy $E_\mathrm{HF}$ is that obtained by using just 
the Slater determinant part of the wavefunction.

\begin{figure}[t]
 \subfloat[Correlation energy missing]{
 \includegraphics[draft=false,width=1.0\linewidth]{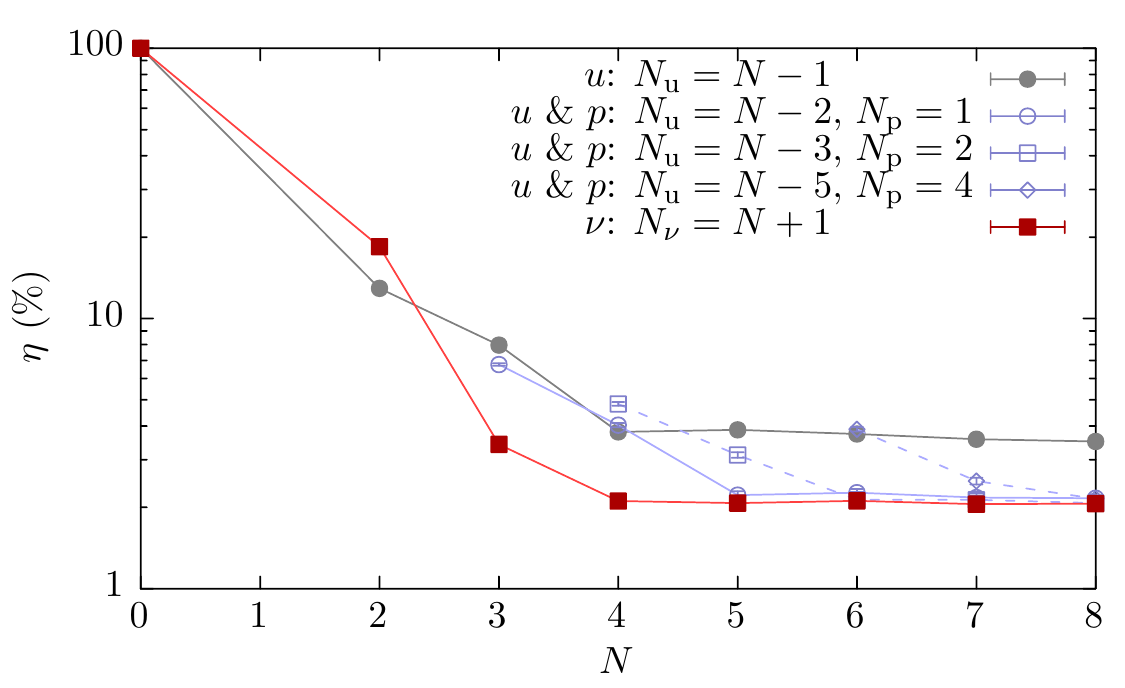}
 \label{fig:HEG_energy}
 }
\\
 \subfloat[Local energy variance]{
 \includegraphics[draft=false,width=1.0\linewidth]{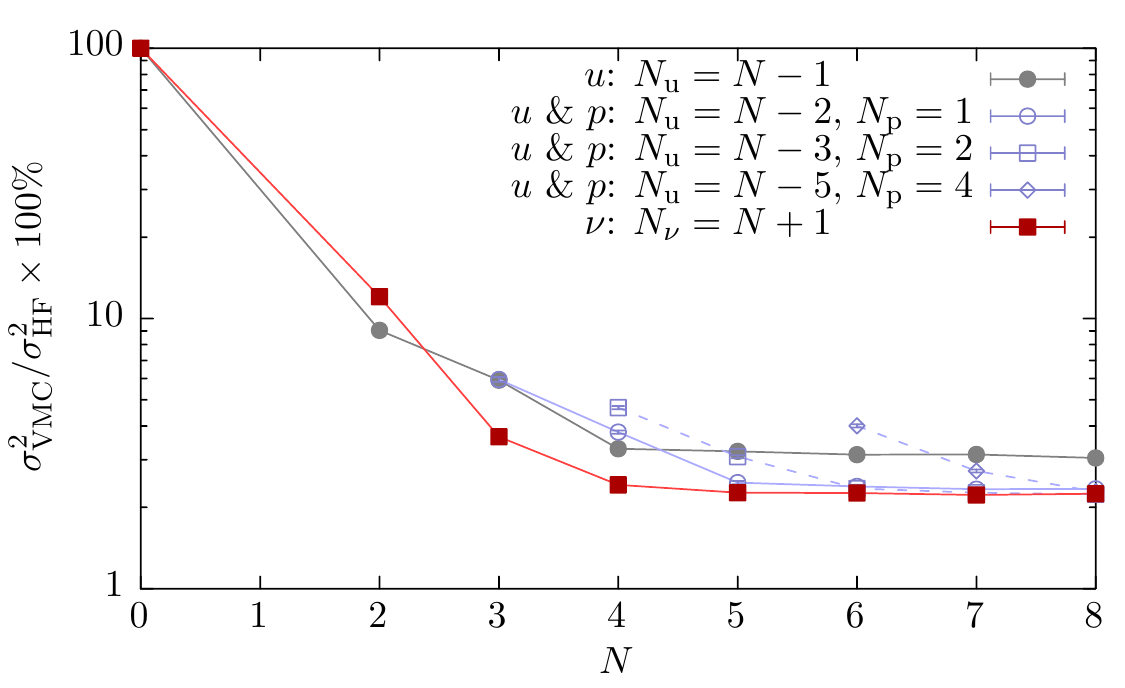}
 \label{fig:HEG_variance}
 }
 \label{fig:HEGenergy}
 \caption{(Color online)
  (a) The percentage of the DMC correlation energy missing from VMC simulations 
of the homogeneous electron gas, with $N$ optimizable parameters in the Jastrow 
factor. Gray, blue, and red lines correspond to the $u$, $u$ \& $p$, and $\nu$ 
term respectively. (b) The variance in the local energy when using different 
Jastrow factors, as a percentage of the variance in the local energy using the 
Hartree--Fock wavefunction.  Error bars, where not visible, are smaller than 
the size of the points.
}
\end{figure}

In \figref{fig:HEG_energy} we compare the percentages of the correlation energy 
missing when the various Jastrow factors under scrutiny are used.  The 
horizontal axis is labeled by the number of optimizable parameters per spin 
channel, $N$, for each Jastrow function: so, for example, a $u$ term with a 
given number $N$ of optimizable parameters per spin channel has 
$N_\mathrm{u}=N-1$ optimizable parameters of terms in the inter-particle 
separation expansion, $\alpha_{m,\sigma\tau}$, as the cutoff length 
$L_{\sigma\tau}$ is also optimized.  A $u$ \& $p$ term with $N_\mathrm{p}$ 
optimizable parameters $a_{\ell,\sigma\tau}$ in the $p$ part leaves 
$N_\mathrm{u}=N-1-N_\mathrm{p}$ optimizable parameters for the $u$ term 
coefficients $\alpha_{m,\sigma\tau}$. For the $\nu$ term $N_\nu=N+1$, as the 
first coefficient $c_{1,\sigma\tau}$ is set by the Kato cusp conditions.  The 
number of optimizable parameters $N$ required to reach a converged accuracy is 
an important measure of the practicality of the Jastrow factors, as $N$ governs 
the complexity of the variance minimization procedure.

We observe that a $u$ term alone can capture over 96\% of the correlation 
energy missing from the Hartree--Fock ($N=0$) result, converging when 
$N_\mathrm{u} \ge 3$ 
($N\ge 4$).  The addition of $p$ terms improves this to only 2\% of 
the correlation energy missing, as inter-particle correlations in the corners 
of the simulation cell are now captured.  The  number of $p$ terms used (if 
greater than zero) makes a negligible difference to the percentage of the 
correlation energy captured, as long as there are also sufficiently many $u$ 
terms present ($N_\mathrm{u}\ge 3$, for a total of $N\ge 5$).  The smallest
number of variational parameters required to achieve convergence is
$N=5$. It is important to 
capture all the short-ranged correlations at the center of the cell, and 
it is also important to capture the leading long-range correlations that 
reflect the symmetry of the simulation cell. This motivates the construction of 
the $\nu$ term as being based around a short-ranged expansion in inter-particle 
separation that interpolates to the lowest-order symmetries of the simulation 
cell at long range.

The $\nu$ Jastrow factor reproduces the best $u$ \& $p$ accuracy of $\eta=2\%$ 
for $N\ge4$.  The need for only $N=4$ optimizable parameters as opposed to the 
$N\ge 5$ required for the $u$ \& $p$ terms means the $\nu$ term is easier to 
optimize. Furthermore, the $\nu$ term has a single parameter $N_\nu$ that can
be increased to improve accuracy, as opposed to having to choose both 
$N_\mathrm{u}$ and $N_\mathrm{p}$ for the $u$ \& $p$ term, which reduces 
the size of the parameter space that needs to be explored.

The $\nu$ term has captured all of the correlation energy available to the $u$ 
\& $p$ terms in this system, but another important quantity in QMC methods is 
the variance in the estimate of the energy.  The variance of the local energy 
determines the efficiency of DMC simulations \cite{Foulkes01,Lee11} and also 
acts as a proxy for the quality of trial wavefunctions, as the variance in the 
local energy of the exact ground state is zero.  In \figref{fig:HEG_variance} 
we examine the variances in the local energy using the different Jastrow 
factors relative to the variance using the Hartree--Fock wavefunction.  Again 
the $u$ term converges for $N\ge 4$, and the addition of $p$ terms reduces the 
variance by another 33\% if a good choice of $N_\mathrm{u}$ and $N_\mathrm{p}$ 
is made with $N\ge 5$.  The $\nu$ term achieves the same reduction in the 
variance in the local energy as these more complicated terms but with fewer 
optimizable parameters, $N\ge 4$.

\begin{figure}
 \includegraphics[width=\linewidth]{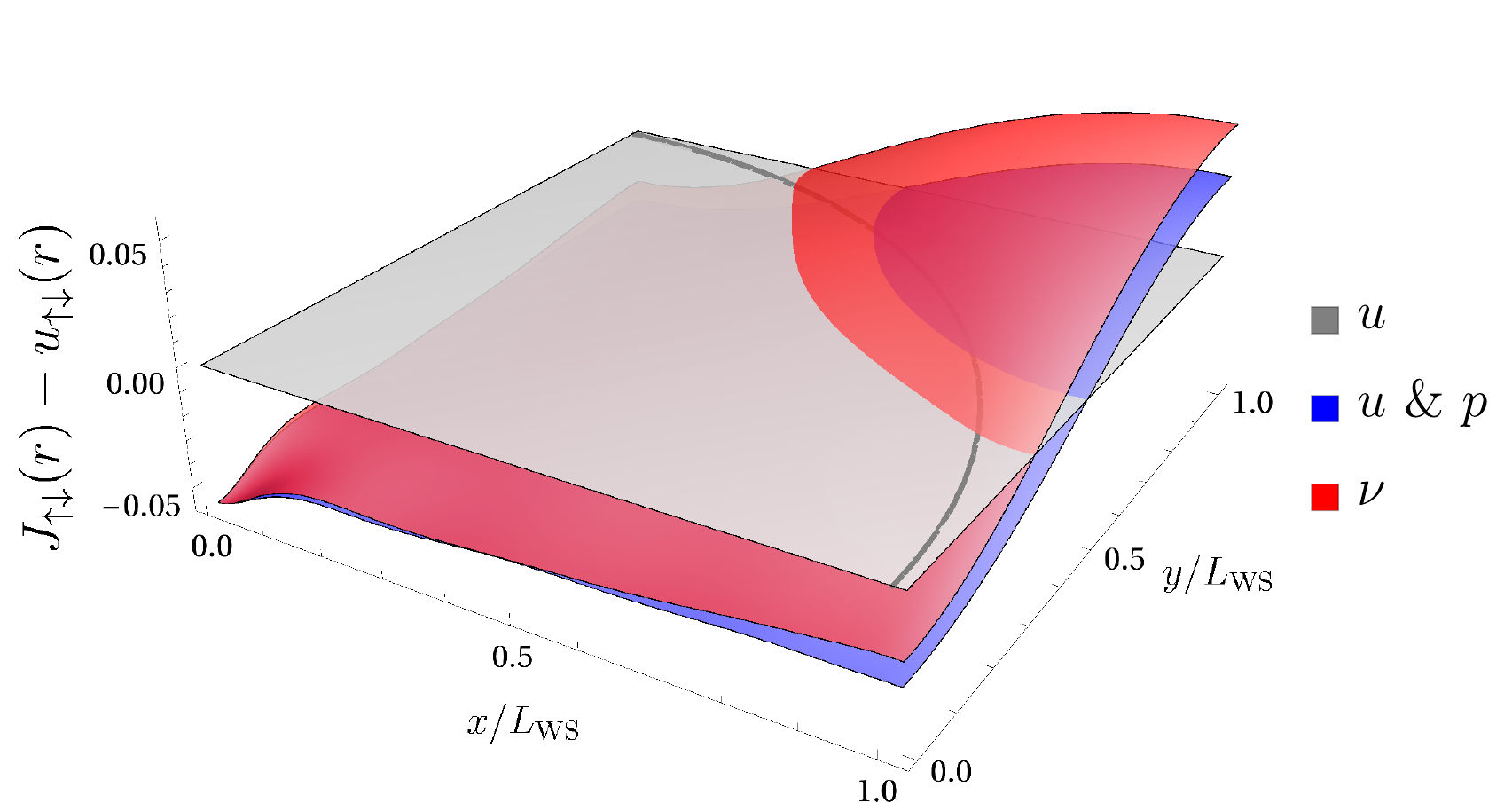}
 \caption{(Color online)
The $N=5$ Jastrow functions with the $u$ Jastrow function subtracted, to show 
how the Jastrow functions vary at large inter-particle separation.
}
 \label{fig:functions_diff}
\end{figure}

The similar levels of the correlation energy captured by the $\nu$ and $u$ \& 
$p$ terms may be understood in terms of the correlations described by these 
Jastrow factors. In \figref{fig:functions_diff} we show the $\nu$ and 
$N_\mathrm{p}=1$ $u$ \& $p$ Jastrow functions with the $u$ Jastrow function 
subtracted, to allow us to focus on the long-range correlations.  Both Jastrow 
functions capture non-trivial correlations in the corner of the simulation 
cell, outside the radius where the $u$ term is cut off to zero (shown by a gray 
arc), explaining their improved performance over the $u$ term.  Furthermore, 
the correlations captured by the $\nu$ and $u$ \& $p$ terms are very similar, 
confirming that both are able to be optimized to capture all of the available 
correlation energy.  The similarity of the $\nu$ and $u$ \& $p$ terms also 
ensures that the zero-wavevector limits of their Fourier transforms are 
likewise similar, 
and hence that the finite-size errors from the Jastrow factors are comparable 
and can be dealt with following the same prescription
\cite{Chiesa06,Drummond08}.  

There is one further advantage to using the $\nu$ term in this system, rather 
than a $u$ \& $p$ term.  The $\nu$ term is a polynomial expansion, like the $u$ 
term, and this makes it quicker to evaluate than the $p$ term with its 
sinusoids from each element of the stars of reciprocal lattice vectors.  For 
$N=5$, where both Jastrow factors have fully converged, the $N_\mathrm{p}=1$ 
$u$ \& $p$ term in the Jastrow factor is 61\% slower to evaluate than the $\nu$ 
term, and the $N_\mathrm{p}=2$ term takes over twice as long to evaluate as the 
$\nu$ term.  This means that simulations with the $\nu$ term can be run 
significantly quicker than those with the $u$ \& $p$ term, to obtain similar 
accuracy.

We have shown that the $\nu$ Jastrow factor captures the ground state energy of 
the HEG as well as a combination of the $u$ and $p$ terms, achieving the same 
accuracy and reduction in variance in the local energy.  In addition to this, 
the $\nu$ term is easier to transfer between systems, as there is only one 
choice of parameter to make as opposed to two for the $u$ \& $p$ term; the 
$\nu$ term requires $N=4$ linear optimizable parameters to converge, rather 
than $N=5$ non-linear parameters for the $u$ \& $p$ term, making it cheaper to 
optimize; and the $\nu$ term is also quicker to evaluate.  We now go on to test 
the $\nu$ Jastrow factor in an inhomogeneous periodic system, for which we take 
the example of crystalline beryllium.

\section{Beryllium}
\label{sec:beryllium}

To demonstrate that the advantages of the $\nu$ term are not restricted to 
simple homogeneous systems with cubic simulation cells, here we test it in a 
crystalline solid.  As discussed in \secref{sec:nu}, the $\nu$ term is 
constructed to interpolate between the symmetry of the interaction potential 
(purely radial) at short radius and the simulation cell symmetry at large 
separation.  In order to demonstrate the generality of this construction, we 
will focus on an analysis of a crystal with relatively low symmetry, in the 
$\mathrm{P}6_3/\mathrm{mmc}$ (hexagonal) space group, where it is non-trivial 
to construct the long-range form of the $\nu$ term.  The simplest example of a 
stable crystal with this space group at zero temperature, where QMC is 
applicable, is crystalline beryllium, and so we use that as our example system. 
 At the end of this Section we will also discuss results in higher-symmetry 
face-centered cubic (FCC) and body-centered cubic (BCC) crystals.

We model crystalline beryllium using an hexagonal simulation cell containing 32 
atoms.  The Be$^{2+}$ ions are represented by pseudopotentials 
\cite{Heine70,Pickett89,Drummond04,Needs10}, and the orbitals in the Slater 
determinants were obtained from a density functional theory 
\cite{Hohenberg64,Kohn65} (DFT) calculation using the {\sc castep} code with a 
plane-wave basis set \cite{Payne92,Clark05}, converted to B-spline functions 
\cite{Hernandez97,Alfe04}.  

The $u$ and $p$ terms are the same in this simulation cell as in the previous 
cubic case, with the $\vect{G}_\ell$ vectors for the $p$ term being the 
reciprocal lattice vectors of the simulation cell, organized into stars of 
equal-length vectors.  In order to use the $\nu$ term we generalize its 
functional form to allow for the use of non-cuboidal simulation cells.

\subsection{Generalized form of \boldmath$\nu$ term}
\label{sec:gennu}

To begin the generalization of the $\nu$ term we construct a set of vectors $\{ 
\vect{B} \}$, formed of the reciprocal lattice vectors of the simulation cell 
and all symmetry-equivalent vectors.  These vectors are exactly those normal to 
the faces of the conventional unit cell, and so encode the symmetry of the 
simulation cell, and have length such that $|\vect{B}_i \cdot 
\vect{r}_\mathrm{face}|=\pi$, for any vector $\vect{r}_\mathrm{face}$ lying in 
the corresponding conventional cell faces.

Constructing a matrix of the reciprocal lattice vectors $\{ \vect{B} \}$ and 
then (left-)inverting and transposing it leads to a set of real-space vectors 
$\{ \vect{A} \}$.  By measuring the projection of the electron-electron 
separation vector $\vect{r}$ onto these real-space vectors we can express the 
electron-electron separation as \mbox{$\vect{r}=\sum_{\zeta} \vect{A}_{\zeta} 
(\vect{B}_{\zeta} \cdot \vect{r})$}.  The inter-particle distance $r$ can then 
be expressed as 
\begin{align*}
r&=\sqrt{\left(\sum_\zeta \vect{A}_\zeta [ \vect{B}_\zeta \cdot \vect{r} 
]\right) \cdot \left(\sum_\xi \vect{A}_\xi [ \vect{B}_\xi \cdot \vect{r} 
]\right)} \neweqnline
&=\sqrt{\sum_i \vect{A}_i \cdot \vect{A}_i w_i^2 + 2\sum_{j>k} \vect{A}_j \cdot 
\vect{A}_k w_j w_k},
\end{align*}
where $w_i=\vect{B}_i \cdot \vect{r}$ expresses the projection of $\vect{r}$ 
onto $\vect{A}_i$ as a phase between $-\pi$ and $\pi$ as $\vect{r}$ runs 
between parallel faces of the conventional cell.  In a directly analogous way 
to the previous, cuboidal form we then define the Jastrow function as
\begin{align}
\nu_{\sigma\tau}(\vect{r})=\sum_{n=1}^{N_\nu} c_{n,\sigma\tau} \Big| &\sum_i 
\vect{A}_i \cdot \vect{A}_i f^2 (w_i) + \neweqnline
& 2\sum_{j>k} \vect{A}_j \cdot \vect{A}_k g(w_j)g(w_k) \Big|^{n/2},
\label{eq:nu}
\end{align}
where in order to reduce to a radial expression at short radius we require that 
$f(w_i)\to |w_i|$ and $g(w_i) \to w_i$ as $\vect{r} \to \boldsymbol{0}$. In 
order to retain the symmetry of the simulation cell at large radii we demand 
$f(w_i)$ be symmetric under $w_i \to -w_i$, whilst $g(w_i)$ is required to be 
antisymmetric, and both functions should satisfy periodic boundary conditions 
at $|w_i|=\pi$.  To satisfy these requirements we take $f$ and $g$ to have the 
simple forms
\begin{align}
f(w_i) &= |w_i|\left( 1 - \frac{|w_i/\pi|^{3}}{4} \right) \neweqnline
g(w_i) &= w_i \left( 1 - \frac{3}{2} |w_i/\pi| + \frac{1}{2} |w_i/\pi|^{2}  
\right).
\label{eq:FandG}
\end{align}
$f(w_i)$ is very similar to the cuboidal form given in \eqnref{eq:cubic_nu}, 
and if we use a cuboidal simulation cell with orthogonal lattice vectors,  
where $\{ \vect{A} \}=\{ \vect{a}_1/2\pi,\vect{a}_2/2\pi,\vect{a}_3/2\pi \}$ 
and $\{ \vect{B} \}=\{ \vect{b}_1,\vect{b}_2,\vect{b}_3 \}$, the general form 
of the Jastrow function \eqnref{eq:nu} reduces to the cuboidal form 
\eqnref{eq:cubic_nu}. $g(w_i)$ is the lowest-order polynomial-like expansion 
that is antisymmetric under $w_i\to -w_i$.  The sets of vectors $\{ \vect{A} 
\}$ and $\{ \vect{B} \}$ that we use for the hexagonal simulation cell, as well 
as for other common simulation cell geometries, are given in the Appendix.

\subsection{Electron-ion correlations}

In crystalline systems there are correlations between the ions and electrons, 
as well as those between electrons.  The DFT orbitals in the Slater 
determinants describe most of the electron-ion correlations, but these are 
modified by the introduction of electron-electron correlations in the Jastrow 
factor: in our simulations we add optimizable electron-ion correlations to the 
electron-electron Jastrow factor to counter this,
\begin{align*}
J(\vect{R}) = \sum_{\substack{j> i\\\sigma,\tau\in\{\uparrow,\downarrow\}}} 
J_{\sigma\tau}(\vecrij) + 
\sum_{\substack{i,I\\\sigma\in\{\uparrow,\downarrow\}}} \chi_\sigma 
(\vect{r}_{iI}),
\end{align*}
where $\vect{r}_{iI}=\vect{r}_i-\vect{r}_I$, for ion positions $\vect{r}_I$, 
$i$ running over all electrons, and $I$ running over all ions.  It has been 
shown \cite{Drummond04,Rios12} that a short-ranged $u$-like expansion in 
electron-ion separation,
\begin{align*}
\chi_\sigma(\vect{r}_{iI})=&\left( \frac{L^\chi_\sigma}{3}\beta_{1,\sigma} + 
\sum_{m=1}^{N_\chi}\beta_{m,\sigma} r_{iI}^m \right)\times\neweqnline
 &\times \left( 1-r_{iI}/L^\chi_\sigma \right)^3 \Theta(L^\chi_\sigma-r_{iI}),
\end{align*}
captures the most important electron-ion correlations in the electron-ion term, 
without the need for a longer-ranged $p$-like term.  The cutoff length 
$L^\chi_\sigma$ is generally comparable to the inter-ionic distance, and we use 
$N_\chi=4$  in our simulations.  As we use pseudopotentials for the ions there 
is no gradient discontinuity in the wavefunction at electron-ion coincidence. 
We also tested a cuspless form of the $\nu$ Jastrow function to capture the 
electron-ion correlations, which agreed with the energies obtained using 
$\chi_\sigma(\vect{r}_{iI})$
to within $10^{-5}$ a.u. with the same number of variational parameters. This
confirms that it is the short-range electron-ion 
correlations that are the most important to capture, and so we shall use the
well-established $\chi_\sigma(\vect{r}_{iI})$ term in the following 
investigations.

In all-electron QMC simulations, particularly of molecules, the addition of 
three-body electron-electron-ion correlations to the Jastrow factor lowers 
the calculated energy \cite{Foulkes01,Rios12,Huang97}, 
as these terms allow a more detailed description of tightly-bound electrons.
However, electron-electron-ion correlations are less important in simulations 
using pseudopotentials, and including them here changes the correlation energy
 by 
less than 0.9\%.  Similarly to the electron-ion term the dominant effect of 
electron-electron-ion terms is at short radius, and so the $\nu$ Jastrow factor
formalism is expected to offer limited improvements relative to an isotropic 
$u$-like term in constructing such terms.  As electron-electron-ion terms make 
a small difference to the energy and will
not help us to discrimitate between the $u$, \mbox{$u$ \& $p$}, and $\nu$ terms
we neglect them here, although they should of course be included in simulations
targeting high accuracy.

The full Jastrow function is
then obtained by combining the electron-ion term $\chi_\sigma(\vect{r}_{iI})$
with the electron-electron Jastrow functions under examination, the $u$, 
\mbox{$u$ \& $p$}, and generalized $\nu$ terms.  We now examine the accuracy
and efficiency of these Jastrow functions for simulating crystalline beryllium.

\subsection{Results}

\begin{figure}[t]
 \subfloat[Correlation energy missing]{
 \includegraphics[draft=false,width=1.0\linewidth]{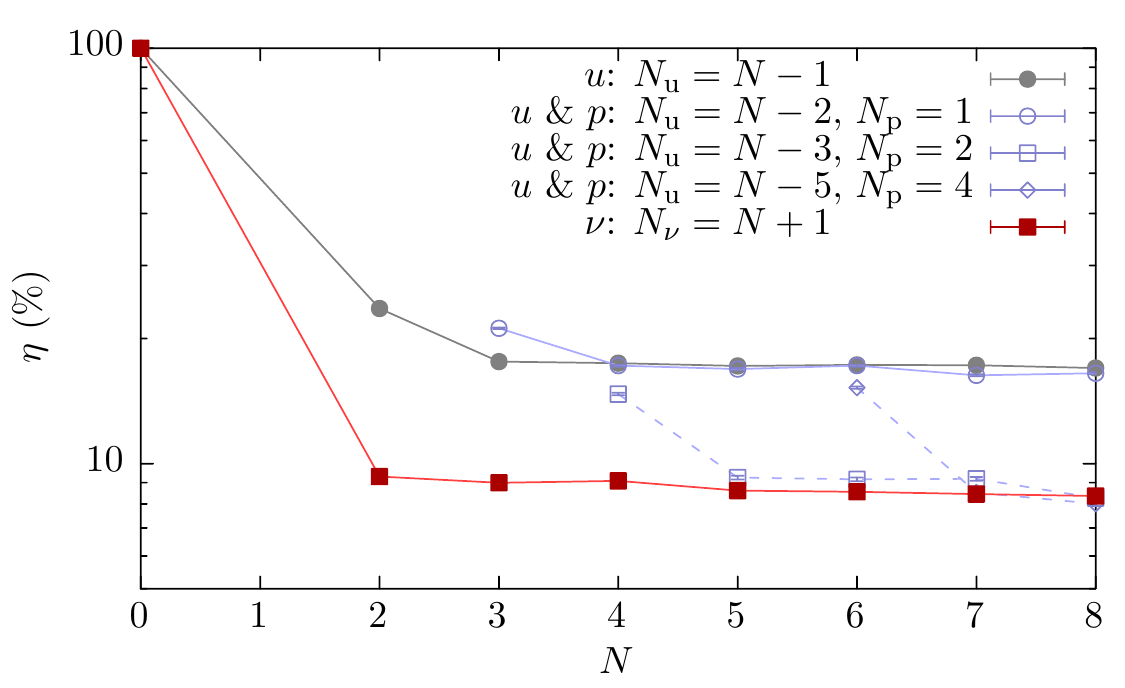}
 \label{fig:Be_energy}
 }
\\
 \subfloat[Local energy variance]{
 \includegraphics[draft=false,width=1.0\linewidth]{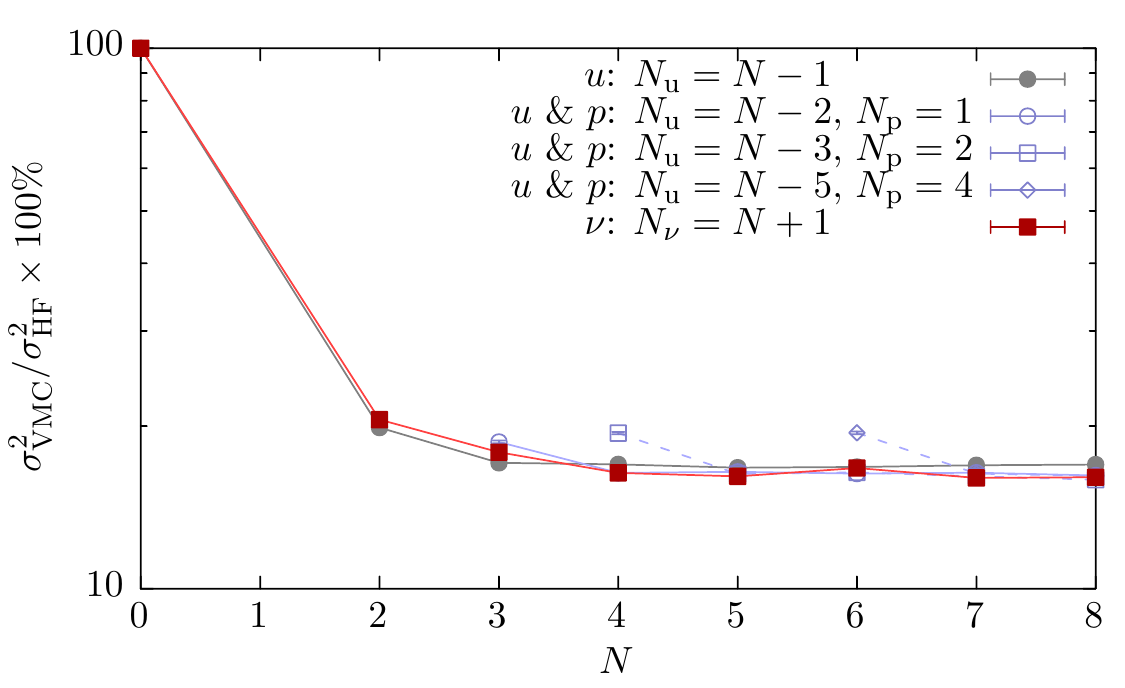}
 \label{fig:Be_variance}
 }
 \label{fig:Beenergy}
 \caption{(Color online)
  (a) The percentage of the DMC correlation energy missing from VMC simulations 
of crystalline beryllium, with $N$ optimizable parameters in the Jastrow 
factor. Gray, blue, and red lines correspond to the $u$, $u$ \& $p$, and $\nu$ 
term respectively. (b) The variance in the local energy when using different 
Jastrow factors, as a percentage of the variance in the local energy using the 
Hartree--Fock wavefunction.  Error bars, where not visible, are smaller than 
the size of the points.
}
\end{figure}

In \figref{fig:Be_energy} we compare the percentages of the DMC correlation 
energy missing, $\eta$, when the various Jastrow factors are used with $N$ 
variational parameters in the electron-electron Jastrow factor.  We observe 
that a $u$ term alone is always missing nearly 20\% of the correlation energy, 
and moreover that the addition of a single $p$ term does not significantly 
improve the result.  This is in contrast to the case of the HEG, where the 
addition of a single $p$ term was the most important step in achieving a 
high-accuracy $u$ \& $p$ term.  This is due to the fact that, in the beryllium 
simulation cell, the $\vect{b}_3$ lattice vector orthogonal to the hexagonal 
planes is shorter than those in the $\vect{b}_1,~\vect{b}_2$ plane, and so the 
first $p$ term only acts along the $c$ axis, not providing flexibility to 
capture correlations in the hexagonal planes.  However, the addition of just 
one more $p$ term reduces the correlation energy missing to around 9\%, and the 
addition of more $p$ terms to this does not significantly alter the result.  
This means that to achieve convergence we again require $N= 5$ when using 
the $u$ \& $p$ term.

As in the HEG, the $\nu$ term achieves comparable accuracy to the most accurate 
$u$ \& $p$ terms, reaching convergence by $N=2$. This, combined with the
necessity of otherwise 
using $N\ge 5$ for the $u$ and $p$ term, of which $N_\mathrm{p}=2$ are 
expensive $p$ terms, means that the $\nu$ term is significantly cheaper to 
optimize and use 
than alternative Jastrow factors.

In \figref{fig:Be_variance} we examine the variance in the local energy using 
different Jastrow factors.  There is significantly less difference here between 
the Jastrow factors than in the proportion of the correlation energy they 
capture, but the $\nu$ term again performs as well as the most detailed other 
Jastrow factors, meaning that the trial wavefunctions have similar efficiency 
in DMC.

\begin{table}[tb]
\begin{ruledtabular}
\begin{tabular}{c . .}
\multicolumn{1}{c}{Crystal type (example system)} & \multicolumn{1}{c}{$\nu$} & 
\multicolumn{1}{c}{$u$ \& $p$: $N_\mathrm{p}=2$} \\
\colrule
Hexagonal (Be) & 8.6(1)\% & 9.3(1)\% \\
BCC (Li) & 4.7(1)\% & 5.0(1)\% \\
FCC (Si) & 10.1(2)\% & 9.6(2)\% \\
\end{tabular}
\end{ruledtabular}
\caption{The percentage of the DMC correlation energy missing within VMC, 
$\eta$, for the $\nu$ and $u$ \& $p$ terms with $N=5$ in example systems: 
crystalline beryllium in a hexagonal simulation cell; crystalline lithium in a 
body-centered cubic (BCC) simulation cell; and crystalline silicon in a 
face-centered cubic (FCC) simulation cell. Bracketed numbers indicate the 
standard error in the values for $\eta$.}
\label{tab:crystals}
\end{table}

As well as hexagonal crystalline beryllium, we have also tested the 
electron-electron $\nu$ term in other crystals with different symmetry.  The 
missing correlation energy when using the $\nu$ term and the $N_\mathrm{p}=2$ 
$u$ \& $p$ term is shown in \tabref{tab:crystals}. The $u$ \& $p$ term is not 
significantly improved by increasing $N_\mathrm{p}$ in any of these crystals, 
and we use $N=5$ as this is where the $u$ \& $p$ term approaches its converged 
accuracy; in each case the $\nu$ term is already converged.

The two Jastrow factors capture similar levels of the correlation energy in 
each system, indicating that the $\nu$ term is a good general choice of Jastrow 
factor for use in crystalline systems, with the slight differences between the 
$\nu$ and $u$ \& $p$ terms in different systems being due to the exact details 
of the symmetry of the simulation cell in each case, some of which are better 
captured by the $\nu$ term than others.  However, overall the differences 
between Jastrow factors are smaller than the differences between systems, and 
the $\nu$ Jastrow factor achieves high accuracy whilst having fewer (and only 
linear) parameters to optimize and being cheaper to evaluate, due to being 
polynomial as opposed to sinusoidal.

\section{Discussion}
\label{sec:Discussion}

We have proposed and tested a form of electron-electron Jastrow factor that 
interpolates between the radial symmetry of the Coulomb potential at short 
range and the space-group symmetry of the simulation cell at large separation.  
The $\nu$ Jastrow factor captures comparable levels of the correlation energy 
to the most detailed $u$ \& $p$ terms used in the literature, and converges 
with fewer variational parameters.  There is also only one choice of input to 
the $\nu$ term, the expansion order $N_\nu$, which reduces the parameter space 
to be explored compared to the two variables, $N_\mathrm{u}$ and 
$N_\mathrm{p}$, required for the $u$ \& $p$ term.  Finally, the polynomial 
$\nu$ term is quicker to evaluate than the plane-wave $p$ term.

It would be possible to apply the ideas behind the $\nu$ term to higher 
angular-momentum terms in a Jastrow factor: for instance, carrying out the 
transformation \mbox{$x/r \to g(x)/\sqrt{f^2(x)+f^2(y)+f^2(z)}$} would allow 
the $Y_{11}$ spherical harmonic to be expressed in a way that satisfies the 
symmetry of a cuboidal simulation cell.  The $\nu$ term could also be used in
systems with interactions other than the Coulomb potential; for instance, QMC 
may also be used to study the dipolar \cite{Whitehead16} and contact 
\cite{Whitehead16a} interactions in cold atomic gases, and also more exotic 
interactions such as those found in 2D semiconductors \cite{Ganchev15}.  The 
interpolation between symmetries 
of the $\nu$ term could also be applicable more widely than just in Jastrow 
factors.  Any expansion in or use of inter-particle separation in a numerical 
investigation could be written instead in terms of the $f$ and $g$ functions of 
the $\nu$ term, and so would immediately satisfy periodic boundary conditions 
in the simulation cell.  Systems that might be well-suited to this approach 
could include two-particle pairing orbitals in Slater determinants 
\cite{Bouchaud88}, large-amplitude phonons simulated within density functional 
theory \cite{Baroni01}, or the construction of force fields that natively 
reflect bond angles for molecular dynamics simulations \cite{Mackerell04}.

The $\nu$ Jastrow factor is implemented in the {\sc casino} QMC package 
\cite{Needs10,casino}.  Data used for this paper are available online 
\cite{repository}.

\begin{acknowledgments}
  The authors thank Pablo L\'opez R\'ios and Neil Drummond for useful 
  discussions,
  and acknowledge the financial support of the
  EPSRC [EP/J017639/1]. G.J.C. also acknowledges the financial support of 
Gonville \& Caius College and the Royal Society.
\end{acknowledgments}

\appendix*
\section{Symmetry-related vectors for the \boldmath{$\nu$} term}

Here we enumerate the $\{ \vect{A} \}$ and $\{ \vect{B} \}$ vectors for use in 
the $\nu$ term for some common simulation-cell geometries.

\subsection{Cubic cell}
For a cubic cell with lattice vectors $\vect{a}_1=a[100]$, $\vect{a}_2=a[010]$, 
$\vect{a}_3=a[001]$, the symmetry-related vectors take the form
\begin{align*}
\{ \vect{A} \}&=\frac{1}{2\pi}\{ \vect{a}_1, \vect{a}_2, \vect{a}_3 
\}\neweqnline
\{ \vect{B} \}&=\{ \vect{b}_1, \vect{b}_2, \vect{b}_3 \}.
\end{align*}

\subsection{FCC cell}
For a face-centered cubic cell with lattice vectors 
$\vect{a}_1=\frac{a}{2}[011]$, $\vect{a}_2=\frac{a}{2}[101]$, 
$\vect{a}_3=\frac{a}{2}[110]$, the symmetry-related vectors take the form
\begin{align*}
\{ \vect{A} \} & = \frac{1}{8\pi} \{ 3\vect{a}_1-\vect{a}_2-\vect{a}_3 , 
3\vect{a}_2 - \vect{a}_3 - \vect{a}_1 , \neweqnline
&\qquad \quad \, 3\vect{a}_3- \vect{a}_1 - \vect{a}_2 , \vect{a}_1 + \vect{a}_2 
+ \vect{a}_3 \} \neweqnline
\{ \vect{B} \} &= \{ \vect{b}_1,\vect{b}_2,\vect{b}_3,\vect{b}_1 + \vect{b}_2 + 
\vect{b}_3 \}.
\end{align*}

\subsection{BCC cell}
For a body-centered cubic cell with lattice vectors 
$\vect{a}_1=\frac{a}{2}[\bar{1}11]$, $\vect{a}_2=\frac{a}{2}[1\bar{1}1]$, 
$\vect{a}_3=\frac{a}{2}[11\bar{1}]$, the symmetry-related vectors take the form
\begin{align*}
\{ \vect{A} \} &= \frac{1}{8\pi} \{ 2 \vect{a}_1 + \vect{a}_2 + \vect{a}_3 , 2 
\vect{a}_2 + \vect{a}_3 + \vect{a}_1 , \neweqnline
&\qquad \quad \, 2 \vect{a}_3 + \vect{a}_1 + \vect{a}_2 , \vect{a}_1 - 
\vect{a}_2 , \vect{a}_1 - \vect{a}_3, \vect{a}_2 - \vect{a}_3 \} \neweqnline
\{ \vect{B} \} &= \{ \vect{b}_1, \vect{b}_2,\vect{b}_3, \vect{b}_1 - 
\vect{b}_2, \vect{b}_1 - \vect{b}_3,\vect{b}_2 - \vect{b}_3 \}.
\end{align*}

\subsection{Hexagonal cell}
For a hexagonal cell with lattice vectors $\vect{a}_1=a[100]$, 
$\vect{a}_2=a[\frac{1}{2} \frac{\sqrt{3}}{2} 0]$, $\vect{a}_3=c[001]$, the 
symmetry-related vectors take the form
\begin{align*}
\{ \vect{A} \} &=\frac{1}{6\pi} \{ 2\vect{a}_1 - \vect{a}_2, 2\vect{a}_2 - 
\vect{a}_1, 3 \vect{a}_3, \vect{a}_1 + \vect{a}_2 \} \neweqnline
\{ \vect{B} \} &=\{ \vect{b}_1, \vect{b}_2, \vect{b}_3, \vect{b}_1 + \vect{b}_2 
\}.
\end{align*}

\end{document}